\documentclass[twocolumn,prb,superscriptaddress]{revtex4}
\usepackage{graphicx}
\usepackage{dcolumn}
\usepackage{bm}
\usepackage{amsmath}
\usepackage{times}
\usepackage{color}
\usepackage[breaklinks=true,colorlinks,citecolor=blue,linkcolor=blue,urlcolor=blue]{hyperref}

\makeatletter

\newcommand{\Rmnum}[1]{\expandafter\@slowromancap\romannumeral #1@}
\makeatother

\begin{document}

\title{Frustration induced inversion of magnetocaloric effect and enhanced cooling power in substituted Pyrochlore Iridates}

\author{Vinod Kumar Dwivedi}
\email{vinodd.iitbombay@gmail.com}
\affiliation{Materials Science Programme, Indian Institute of Technology Kanpur, Kanpur 208016, India}
\thanks{Present Address: Department of Physics, Indian Institute of Technology Bombay, Powai, Mumbai 400076, India}

\author{Prabhat Mandal}
\affiliation{Saha Institute of Nuclear Physics, HBNI, 1/AF Bidhannagar, Kolkata 700064, India}

\author{Soumik Mukhopadhyay}
\email{soumikm@iitk.ac.in}
\affiliation{Department of Physics, Indian Institute of Technology Kanpur, Kanpur 208016, India}


\begin{abstract}
We investigate the effect of partial replacement of extended $5d$ Ir$^{4+}$ sites by localized $3$d Cr$^{3+}$ moments on the magnetocaloric properties of Y$_2$Ir$_2$O$_7$ (YIO) pyrochlore iridates. We find that Y$_2$Ir$_{2-x}$Cr$_x$O$_7$ (YICO) undergoes cluster glass transition, possibly due to RKKY like interaction between localized Cr$^{3+}$ moments occupying random sites in the pyrochlore network, mediated by $5d$ Ir conduction electrons. The coexistence of ferromagnetic and antiferromagnetic clusters give rise to the conventional and inverse magnetocaloric effect (MCE). We observe significant enhancement of conventional as well as inverse MCE with substitution. Although the value of conventional MCE and inverse MCE in substituted Iridates are not large, the effect spans over a giant working temperature window, thus leading to orders of magnitude enhancement of cooling power, the value being comparable to standard magnetocalric materials.

\end{abstract}

\maketitle
In recent years, solid state magnetic cooling based on the magnetocaloric effect (MCE) has gained considerable attention as an alternative to the traditional gas compression refrigeration technology~\cite{Julia,Jian,Pecharsky1}. The MCE is defined and measured using primarily two broad categories: 1) change in the temperature of magnetic material due to the application of magnetic field in an adiabatic process ($\bigtriangleup T_{ad}$); and 2) change in the magnetic entropy ($\bigtriangleup S_M$) with application of magnetic field in an isothermal condition. In conventional magnetocaloric effect, spins align on application of magnetic field leading to reduction of magnetic entropy: the materials heat up when the magnetic field is applied and cool down when the field is removed. However, the opposite is also possible when the magnetic entropy increases with application of magnetic field. Such phenomenon is characterized as inverse MCE (IMCE)~\cite{Krenke,Xixiang,Anis2,Ajay,QZhang}. The IMCE materials exhibit minimum in $-\bigtriangleup S_M$ vs $T$ curve near AFM transition temperature ($T_N$). Generally, systems having AFM, ferrimagnetic transitions and ferromagnetic materials with martensitic phase transformation, are known to exhibit IMCE~\cite{Krenke,Anis2}. IMCE materials can be used for cooling by application of magnetic field under adiabatic condition and can also be employed as heat-sinks in case of cooling by conventional MCE materials~\cite{Krenke,Anis2}. Thus the discovery of new IMCE materials is equally important, if not more, in the search for suitable conventional MCE materials.

The refrigerant capacity (RC) provides an estimate of the amount of transferred heat between the hot end at $T_{hot}$ and cold end at $T_{cold}$ in one ideal thermodynamic cycle~\cite{QZhang}. While comparing the different materials to be utilized in the same thermodynamic cycle, the materials exhibiting larger RC is favored because of larger amount of heat transfer. RC is defined as the area under the $-\bigtriangleup S_M(T)$ curve for a particular magnetic field between $T_{hot}$ and $T_{cold}$, i.e., RC$=\int^{T_{hot}}_{T_{cold}}[\bigtriangleup S_M(T)]dT$. The two temperatures $T_{cold}$ and $T_{hot}$ define the working range of the refrigerator, which is associated with the full width at half maximum ($\delta T_{FWHM}$) of the $-\bigtriangleup S_M(T)$ curve.

It was proposed long back that the strongly frustrated classical Heisenberg antiferromagnets could potentially exhibit greater field induced adiabatic temperature change as compared to non-frustrated magnets~\cite{Zhitomirsky}. However, experimental reports of MCE in the frustrated pyrochlore oxides with formula A$_2$B$_2$O$_7$ (A=Y, Bi, rare earth elements; and B$=$ transition metal) are not very encouraging and are limited to Mn~\cite{Cai,Yikun,Cui,Ben1,Khachnaoui}, Ti~\cite{Aoki,Orendac,Sosin,Ben3,Wolf}, Mo~\cite{YaoDong} and Sn~\cite{Tkac} based pyrochlore oxides. Investigation of MCE on pyrochlore iridates could be interesting not only from applications point of view but also from the fundamental point of view. Pyrochore iridates provide a vast template to study the interplay of magnetic frustration, spin-orbit coupling and Coulomb correlation~\cite{Krempa,Abhishek1,Bikash,Vinod1,Vinod2,Vinod3,Vinod4,Vinod5}. The resulting complex magnetic phases can thus be probed using MCE as well. In the present work, we investigate the MCE of frustrated pyrochlore iridates Y$_2$Ir$_{2-x}$Cr$_x$O$_7$ (YICO). We discover the coexistence and enhancement of the conventional MCE and IMCE with chemical substitution, accompanied by high refrigerant capacity and relative cooling power with a broad working range around the liquid nitrogen temperature. The present study firmly places YICO in the category of new promising oxides candidates for magnetic cooling.



\begin{figure}
	\centering
	\includegraphics[width=\linewidth]{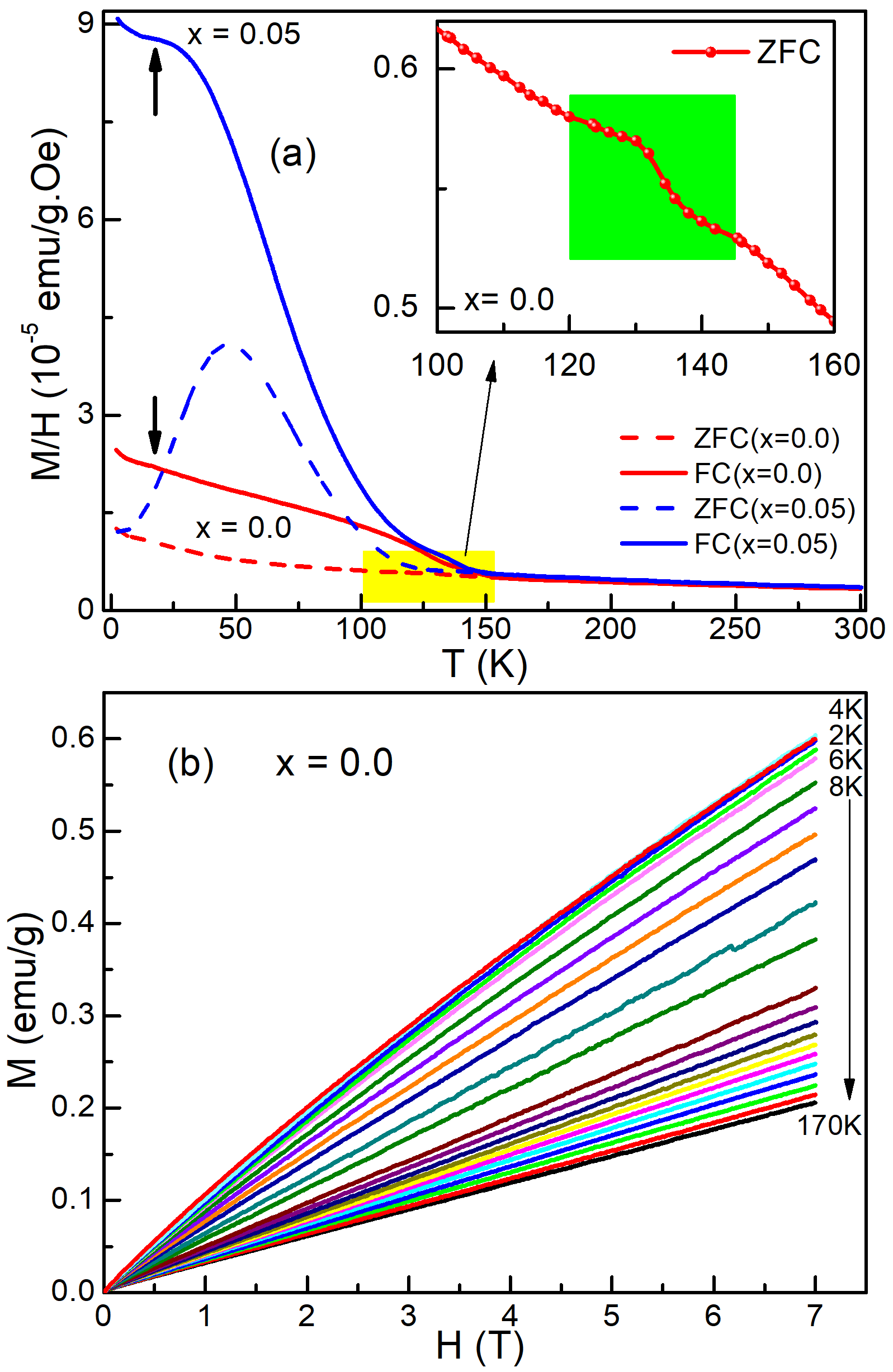}\\
	\caption{(a) Magnetization as a function of temperature; inset shows enlarged view of $\chi_{ZFC}$-T of un-doped sample. The arrows indicate magnetic transitions labeled as $T_N$ in the text. (b) Magnetic field dependence of isotherm magnetization measured at different temperatures for $x =0.0$.}\label{fig:MtMh}
\end{figure}

\begin{figure}
	\centering
	\includegraphics[width=\linewidth]{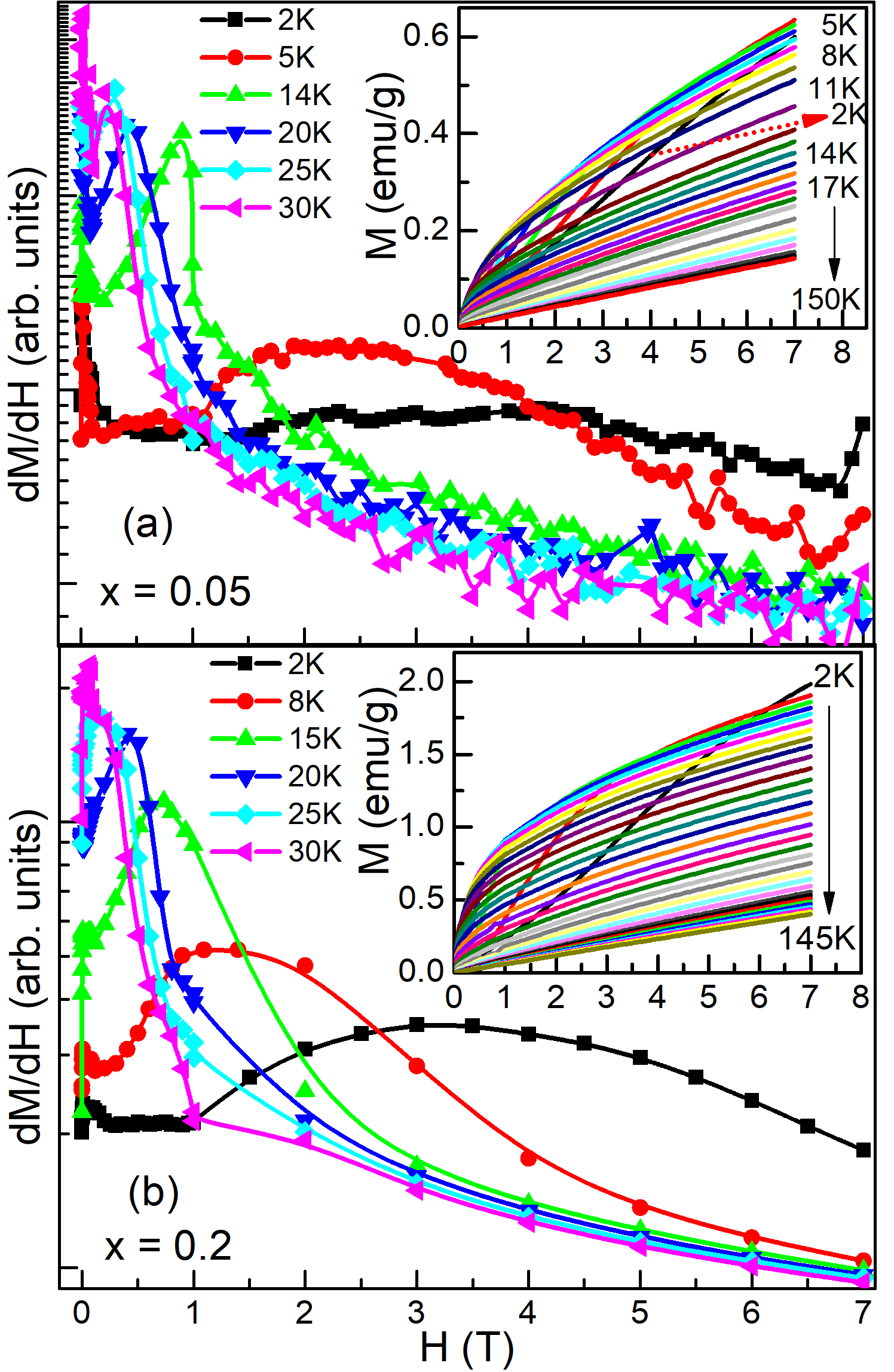}\\
	\caption{First derivative of magnetization as a function of magnetic field for (a) $x=0.05$ and (b) $x=0.2$; peaks represent critical fields for meta-magnetic transitions. Insets show the corresponding magnetization isotherms.}\label{fig:Mh}
\end{figure}

\begin{figure*}
	\centering
	\includegraphics[width=\linewidth]{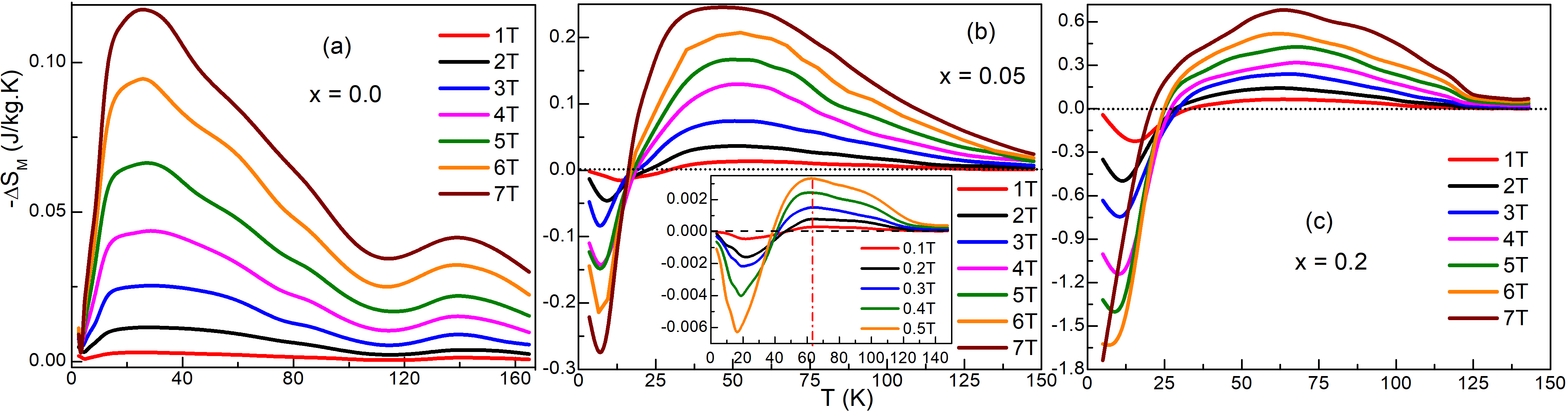}\\
	\caption{Temperature dependence of the magnetic entropy change $-\bigtriangleup S_M$ for the samples (a) $x =0.0$, (b) $x =0.05$; inset shows $-\bigtriangleup S_M(T)$ at low fields and (c) x = 0.2.}\label{fig:Entropy}
\end{figure*}

The synthesis of polycrystalline YICO series with $x = 0.0, 0.05, 0.2$ were done by conventional solid state reaction~\cite{Vinod1}. The raw ingredients of high purity IrO$_2$ (99.99\%), Cr$_2$O$_3$ (99.99\%) and Y$_2$O$_3$ (99.99\%) were used. Mixture of raw materials was ground, pelletized and then sintered at $1000^0$C for $250$~h with several intermediate grindings. The room temperature (Cu-K$_\alpha$ radiation) X-ray diffraction pattern of polycrystalline YICO series shows a nearly pure phase with a cubic $Fd\bar{3}m$ structure~\cite{Vinod1}. Electronic structure characterization was performed using x-ray photoelectron spectroscopy~\cite{Vinod1}. Magnetic measurements were performed using Quantum Design MPMS SQUID VSM.

The field cooled (FC) and zero field cooled (ZFC) magnetic susceptibility $\chi=M/H$ measured as a function of temperature in the presence of applied magnetic field $H =1$~kOe is shown in Fig.~\ref{fig:MtMh}a. For $x=0.0$ sample, a bifurcation can be seen at the irreversibility temperature $T_{irr}$ $\sim158$~K. A maximum in the $\chi_{ZFC}(T)$ curve identified as freezing temperature $T_f$ can be seen for the substituted samples. The spin freezing is not as prominent in the parent undoped sample where $T_{irr}$ and $T_f$ are close to each other (inset Fig.~\ref{fig:MtMh}a) and there is an upturn in ZFC susceptibility below $T_{f}$, indicating only partial spin freezing at $T_f$. The values of $T_f$ for all the samples are given in Table~\ref{T1}. The existence of $T_{irr}$ at higher temperature and maximum in the $\chi_{ZFC}(T)$ curve at lower temperature suggests the possibility of cluster glass-like phase~\cite{Vinod1} particularly in the substituted samples. We estimated the Curie-Weiss temperature $\theta_{CW}$ by fitting the $\chi_{FC}-T$ curve in the high temperature range. The values of $\theta_{CW}$ are found to be negative and large for YICO series, indicating strong effective AFM correlation. The first derivative of $\chi_{ZFC}$ as a function of temperature reveals extremum points, which are identified as FM-like magnetic transition at higher temperature $T_C$~\cite{Ding} and AFM like transition at lower temperature $T_N$~\cite{Kazak}, respectively. The estimated values of $T_C$ and $T_N$ are shown in Table~\ref{T1}.

\begin{figure}
	\centering
	\includegraphics[width=7cm]{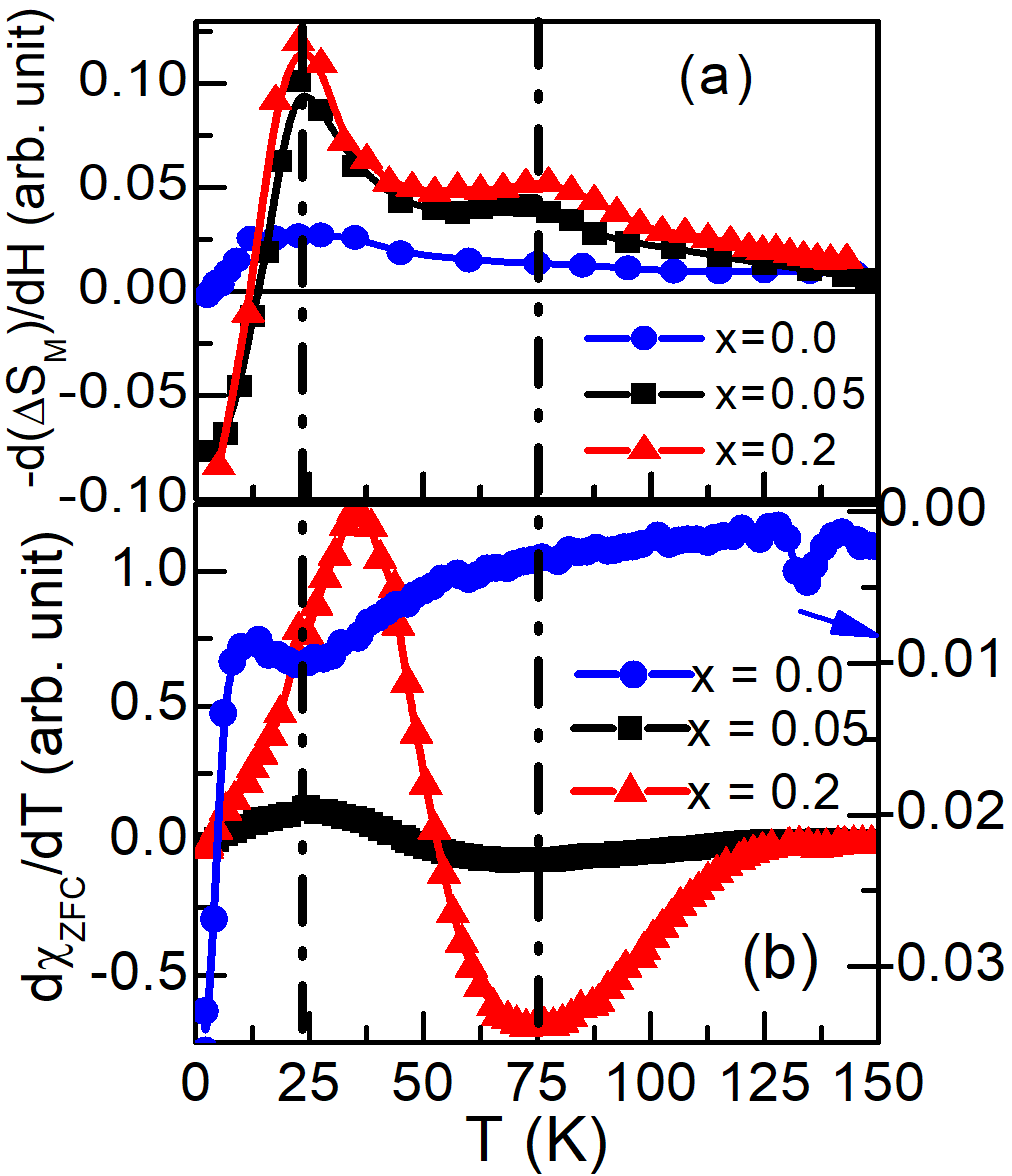}\\
	\caption{(a) Temperature variation of $-\frac{d(\bigtriangleup S_M)}{dH}$ at an applied field
$\lesssim 7$~T. (b) First derivative of $\chi_{ZFC}$ as a function of temperature plotted in the same scale.}\label{fig:Zhitormisky}
\end{figure}

The magnetization isotherms for all the samples have been measured at different closely spaced temperatures. Figure~\ref{fig:MtMh}b and Fig.~\ref{fig:Mh}a-b show that magnetization increases with field up to $7$~T without any sign of saturation. For the doped samples, below $T_N$, the magnetization exhibits an inflection point suggesting metamagnetic like transition further confirmed by the maximum in the slope of $M(H)$ curves (Fig.~\ref{fig:Mh}a,b). The critical field for metamagnetic transition labeled $H_{MT}$ is progressively reduced at elevated temperatures consistent with physical expectations~\cite{Midya}. Although, $H_{MT}$ usually vanishes at $T_N$ for the materials exhibiting PM-AFM transition~\cite{Midya}, it does not vanish completely at $T_N$ for the YICO series, suggesting a more complex scenario such as existence of AFM clusters beyond $T_N$. The high value of $H_{MT}$ suggests presence of stronger magnetic anisotropy in the substituted samples. For the $x=0.0$ sample, the maximum is not observed in $\frac{dM}{dH}$ vs $H$ curves. The metamagnetic transition in substituted samples is more likely of spin-reorientation type rather than the spin-flop type because of small value of $M$ ($0.23 \mu_B$ for $x = 0.2$ at $7$~T and $2$~K).

The change in magnetic entropy $\bigtriangleup S_M$ is calculated from the M-H isotherms using the Maxwell relation~\cite{Ding} $\Delta S_M=\int^H_0(\partial M/\partial T)dH$. Fig.~\ref{fig:Entropy} shows $-\bigtriangleup S_M(T)$ curves plotted for different magnetic fields. For $x=0$, $-\bigtriangleup S_M(T)$ exhibits positive maximum around $T_C$ (Fig.~\ref{fig:Entropy}a). With substitution, $-\bigtriangleup S_M(T)$ shows not just a positive maximum at high temperature, but undergoes sign inversion at $T_f$, followed by a minimum at lower temperature, thus exhibiting the coexistence of conventional MCE and IMCE (Fig.~\ref{fig:Entropy}b,c). Such a smooth cross-over from conventional MCE at high temperature to IMCE at low temperature clearly suggests coexistence and competition between AFM and FM phases. Curiously, for sufficiently low magnetic field, the sign inversion takes place close to $T_f$, presumably a cluster glass freezing transition in presence of competing FM and AFM interaction. The cluster glass behavior cannot be attributed to the geometry of the pyrochlore lattice. It is more likely due to RKKY interaction between local Cr$^{3+}$ moments mediated by Ir$^{4+}$ conduction electrons~\cite{Bikash}.

Further, we have investigated the behavior of MCE near the maximum applied magnetic field of $7$~T, presumably close to the saturation field. Figure~\ref{fig:Zhitormisky}a shows the field derivative of magnetic entropy as a function of temperature, plotted along with $\frac{d\chi_{ZFC}}{dT}$ curve in the Fig.~\ref{fig:Zhitormisky}b. For undoped sample, the value of $-\frac{d(\bigtriangleup S_M)}{dH}$ is positive throughout the temperature range, while the same for doped samples undergoes sign inversion below $T_f$. The enhanced value of $-\frac{d(\bigtriangleup S_M)}{dH}$ and the sign inversion for doped samples could be attributed to the increased frustration due to Cr substitution. The origin of frustration is not purely geometric for which we expect a power law temperature dependence in the field derivative of magnetic entropy~\cite{Zhitomirsky}. We attribute the increased frustration to the competing FM and AFM interactions due to the substitution by localized moment Cr$^{3+}$. The estimated value of frustration parameter $f=\frac{\mid \theta_{CW} \mid}{T_f}$ is listed in Table~\ref{T1}.

\begin{figure*}
	\centering
	\includegraphics[width=12 cm]{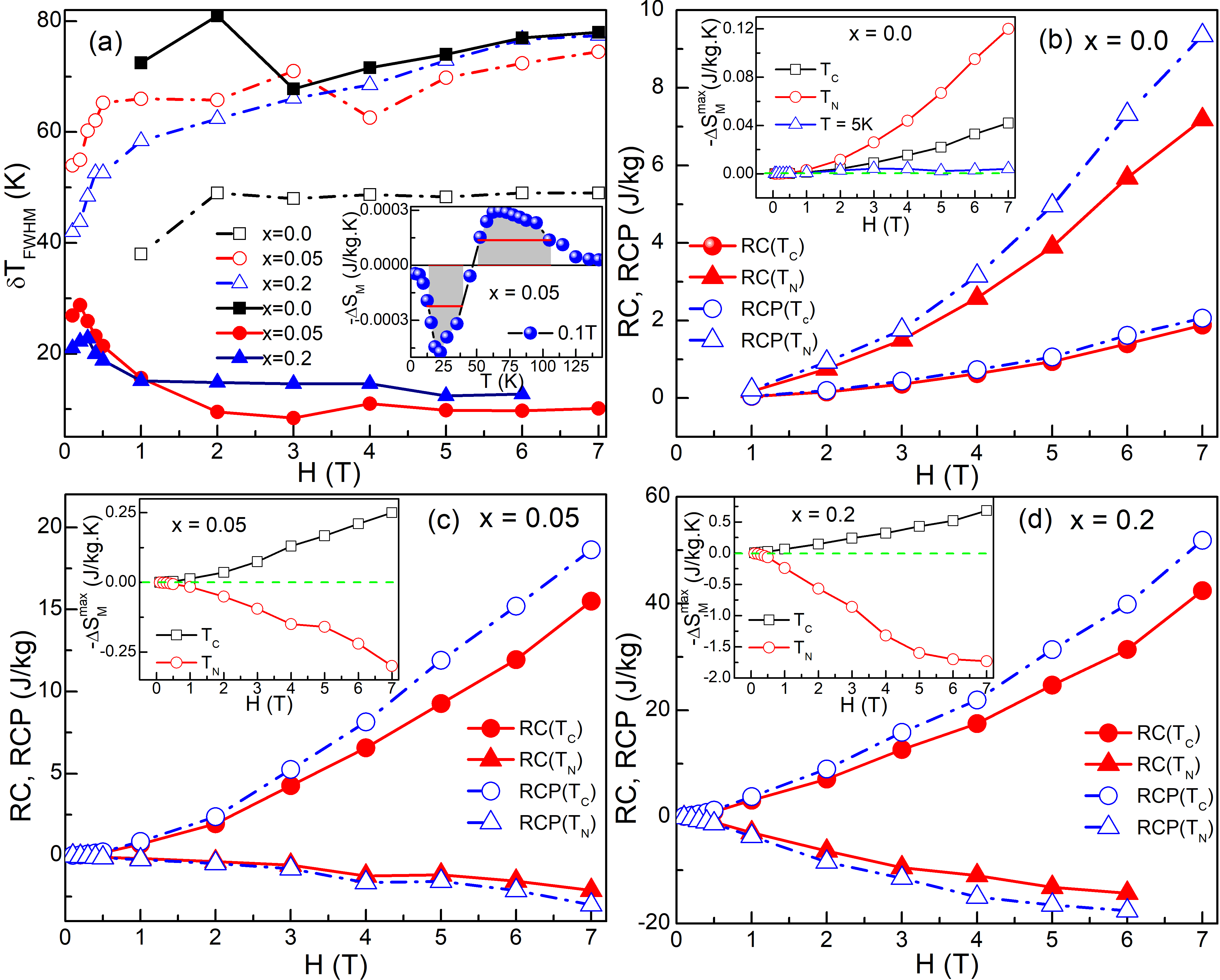}\\
	\caption{(a) Magnetic field dependent $\delta T_{FWHM}$ around $T_C$ [hollow symbol] and $T_N$ [solid symbol]; inset shows the method used to calculate the RC and RCP from the $-\bigtriangleup S_M(T)$ curves around $T_C$ and $T_N$. RC and RCP as a function of applied magnetic fields for the samples (b) $x=0.0$, (c) $x=0.05$, and (d) $x=0.2$. Inset shows $-\bigtriangleup S_M^{max}$ vs H.}\label{fig:RC}
\end{figure*}

\begin{table*}
	\caption{Important parameters related to the magnetocaloric effect. Here, T1 and T2 represent the $\delta T_{FWHM}$ around $T_C$ and $T_N$, respectively. RCP1 and RCP2 are the values of Relative Cooling Power corresponding to T1 and T2, respectively at $7$~T. Curie-Weiss temperature $\theta_{CW}$ is estimated in the temperature range 180-240~K. \label{T1}}
\begin{center}
\begin{tabular}{c c c c c c c c c c c c c}
\hline
$x$&$T_C$&($-\bigtriangleup S_M^{max}$)$_{T_C}$&T1&RCP1&$T_N$&($-\bigtriangleup S_M^{max}$)$_{T_N}$&T2& RCP2& $T_f$& -$\theta_{CW}$& $f$ \\
&K& ($J/Kg.K$) &  ($K$) & ($J/kg$) &($K$) & ($J/Kg.K$) &($K$) &($J/kg$) & ($K$) & (K) &  \\
\hline
0.0  &130&0.042&49&2.1&16&0.12&81&9.4&130 &138& 1.2  \\
0.05 &65&0.25&75&19&20&-0.3&29&-3&46 &155& 2.5  \\
0.2  &70&0.68&78&52&22&-1.73&23&-17.6&50 &140& 3.0 \\
\hline
\end{tabular}
\end{center}
\end{table*}

The values of $-\bigtriangleup S_M^{max}$ are given in Table ~\ref{T1}. We observe a striking enhancement of MCE in substituted samples. However the value still remains small compared to standard magnetocaloric materials. We have calculated RC using the method as shown in the inset of Fig.~\ref{fig:RC}a. The shaded area under the $-\bigtriangleup S_M(T)$ curve is a measure of the RC. For conventional MCE at higher temperature, the value of $\delta T_{FWHM}$ increases with field and reaches up to $81$~K for $7$~T. On the other hand, for IMCE observed at lower temperature, the value of $\delta T_{FWHM}$ decreases with increasing field. The variation of $\delta T_{FWHM}$ as a function of field around $T_C$ and $T_N$ is shown in the Fig.~\ref{fig:RC}a. The calculated values of RC as a function of applied magnetic field around $T_C$ and $T_N$ are shown in Fig.~\ref{fig:RC}b-d. As the formula suggests, either a large value of $-\bigtriangleup S_M(T)$ or a broad $-\bigtriangleup S_M(T)$ curve or both produces large value of RC. In our case the small value of conventional MCE is more than compensated by the large value of $\delta T_{FWHM}$ leading to RC value comparable to standard magnetocaloric materials.

Although the values of $-\bigtriangleup S_M$ for the YICO series are similar to known pyrochlore oxides exhibiting MCE~\cite{Cai,Yikun,Cui,Ben1,Khachnaoui,Aoki,Orendac,Sosin,Ben3,Wolf,YaoDong,Tkac} and not as large as other MCE materials~\cite{Das,Paromita,Sanjay,Santanu1,Santanu2,Chaudhary}, the order of magnitude enhancement of both conventional MCE and IMCE with substitution is quite encouraging. Besides these observations, the relatively wide $\delta T_{FWHM}$ for the YICO samples offers a very competitive RC around both transition temperatures. The coexistence of conventional MCE and IMCE has been observed in a variety of magnetic systems. However, the two working temperature windows are usually small~\cite{Krenke,Xixiang,Anis2,Ajay,Bonilla}. From applications point of view, those magnetocaloric materials which can work over a broad temperature regime i.e, having large $\delta T_{FWHM}$, are highly attractive.

The important findings for the YICO series are that both the `low temperature negative' minimum (around $T_N$) and `high temperatures positive' maximum (around $T_C$) in $-\bigtriangleup S_M(T)$ span over a broad temperatures range. It should also be noted that the reported pyrochlore oxides~\cite{Cai,Yikun,Cui,Ben1,Khachnaoui,Aoki,Orendac,Sosin,Ben3,Wolf,YaoDong,Tkac} shows only conventional MCE and complete absence of IMCE. The value of $\delta T_{FWHM}$ is more than $40$~K around $T_C$ (Fig.~\ref{fig:RC}a) at low field ($0.1$~T for doped sample), which is larger than that of the established reference material Gd~\cite{Julia} but smaller than the recently reported value for Fe-Ni-Cr~\cite{Chaudhary} system. We have also calculated the relative cooling power (RCP) which is often used to estimate their potential for magnetic cooling and defined as RCP$=-\bigtriangleup S_M^{max}\delta T_{FWHM}$. The values of RCP and RC are generally close to each other with the former roughly scaling as $4/3$ times the value of RC. As an example, the value of RCP corresponding to conventional MCE for $x=0.2$ at $5$~T around liquid nitrogen temperature is $0.27$ Jcm$^{-3}$, quite competitive when pitted against known magnetocaloric materials~\cite{Tapas}. Although the IMCE is associated with metamagnetic transition, the hysteresis is negligible, thus minimizing possible energy loss. The RCP corresponding to IMCE being of similar value ($-0.14$ Jcm$^{-3}$ at $5$~T for $x=0.2$), we emphasize that the global working efficiency can be improved further by using both the conventional MCE and IMCE via a special working procedure employing magnetization and demagnetization processes~\cite{Krenke,Xixiang}. Additionally, since it is not based on rare earth elements, YICO might turn out to be an interesting candidate material for magnetic refrigeration.

In conclusion, we find that the Y$_2$Ir$_{2-x}$Cr$_x$O$_7$ (YICO) exhibit multiple magnetic transitions, possibly due to frustration induced coexistence of AFM and FM clusters, which lead to the conventional MCE at high temperature and IMCE at low temperature. One of the most important results of the present work is a plausible strategy whereby magnetic frustration introduced by chemical substitution can lead to not just coexistence of conventional and inverse MCE but order of magnitude enhancement of $-\bigtriangleup S_M$. Both the conventional MCE and IMCE span over a broad working temperature range in YICO, resulting in high values of RC and RCP, comparable to leading magnetocaloric materials. We expect that the present study may open up a pathway to explore more suitable candidate materials among non-rare-earth frustrated magnetic oxides.

\section{ACKNOWLEDGEMENTS}

SM would like to acknowledge Department of Science and Technology (DST), Government of India for financial support.

\section{AUTHOR DECLARATIONS}

\subsection{Data availability}
The data that support the findings of this study are available from the corresponding author upon reasonable request.

\subsection{Conflict of interest}
The authors have no conflicts to disclose.


\begin{thebibliography}{99}
	

\bibitem{Julia} J. Lyubina,
\href{https://doi.org/10.1088/1361-6463/50/5/053002}{{J. Phys. D: Appl. Phys.} {\bf{50}}, 053002 (2017)}.


\bibitem{Jian} J. Liu, T. Gottschall, K. P. Skokov, J. D. Moore and O. Gutfleisch,
\href{https://doi.org/10.1038/nmat3334}{{Nat. Mater.} {\bf{11}}, 620 (2012)}.

\bibitem{Pecharsky1} V. K. Pecharsky and K. A. Gschneidner Jr.,
\href{https://doi.org/10.1103/PhysRevLett.78.4494}{{Phys. Rev. Lett.} {\bf{78}}, 4494 (1997)}.


\bibitem{Krenke} T. Krenke, E. Duman, M. Acet, E. F. Wassermann, X. Moya, L. Manosa, and A. Planes,
\href{https://doi.org/10.1038/nmat1395}{{Nat. Mater.} {\bf{4}}, 450 (2005)}.

\bibitem{Xixiang} X. Zhang, B. Zhang, S. Yu, Z. Liu, W. Xu, G. Liu, J. Chen, Z. Cao, and G. Wu,
\href{https://doi.org/10.1103/PhysRevB.76.132403}{{Phys. Rev. B} {\bf{76}}, 132403 (2007)}.

\bibitem{Anis2} A. Biswas, S. Chandra, T. Samanta, B. Ghosh, S. Datta, M. H. Phan, A. K. Raychaudhuri, I. Das, and H. Srikanth,
\href{https://doi.org/10.1103/PhysRevB.87.134420}{{Phys. Rev. B} {\bf{87}}, 134420 (2013)}.

\bibitem{Ajay} A. Kumar and R. S. Dhaka,
\href{https://doi.org/10.1103/PhysRevB.101.094434}{{Phys. Rev. B} {\bf{101}}, 094434 (2020)}.

\bibitem{QZhang} Q. Zhang, F. Guillou, A. Wahl, Y. Breard, and V. Hardy,
\href{https://doi.org/10.1063/1.3453657}{{Appl. Phys. Lett.} {\bf{96}}, 242506 (2010)}.




\bibitem{Zhitomirsky} M. E. Zhitomirsky,
\href{https://doi.org/10.1103/PhysRevB.67.104421}{{Phys. Rev. B} {\bf{67}}, 104421 (2003)}.


\bibitem{Cai} Y. Q. Cai, Y. Y. Jiao, Q. Cui, J. W. Cai, Y. Li, B. S. Wang, M. T. Fernandez-Diaz, M. A. McGuire, J.-Q. Yan, J. A. Alonso, and J.-G. Cheng, 
\href{https://doi.org/10.1103/PhysRevMaterials.1.064408}{{Phys. Rev. Materials} {\bf{1}}, 064408 (2017)}.

\bibitem{Yikun} Y. Zhang, H. Li, D. Guo, Z. Ren, and G. Wilde,
\href{https://doi.org/10.1016/j.ceramint.2018.05.239}{{Ceram. Int.} {\bf{44}}, 15681 (2018)}.

\bibitem{Cui} Q. Cui, N. N. Wang, N. Su, Y. Q. Cai, B. S. Wang, T. Shinmei, T. Irifune, Jose A. Alonso, and J. G. Cheng,
\href{https://doi.org/10.1016/j.jmmm.2019.165494}{{J. Magn. Magn. Mater.} {\bf{490}}, 165494 (2019)}.

\bibitem{Ben1} N. Ben Amor, M. Bejar, E. Dhahri, M. A. Valente, J. L. Garden, E. K. Hlil,
\href{https://doi.org/10.1007/s10948-013-2188-2}{{J. Supercond. Nov. Magn.} {\bf{26}}, 3455 (2013)}.
\href{https://doi.org/10.1016/j.jallcom.2013.02.008}{{J. Alloys Compd.} {\bf{563}}, 28 (2013)}.


\bibitem{Khachnaoui} F. Khachnaoui, N. Ben Amor, M. Bejar, E. Dhahri, E. K. Hlil,
\href{https://doi.org/10.1007/s10948-018-4656-1}{{J. Supercond. Nov. Magn.} {\bf{31}}, 3803 (2018)}.


\bibitem{Aoki} H. Aoki, T. Sakakibara, K. Matsuhira, and Z. Hiroi,
\href{https://doi.org/10.1143/JPSJ.73.2851}{{J. Phys. Soc. Jpn.} {\bf{73}}, 2851 (2004)}.

\bibitem{Orendac} M. Orendac, J. Hanko, E. Cizmar, A. Orendacova, M. Shirai, and S. T. Bramwell,
\href{https://doi.org/10.1103/PhysRevB.75.104425}{{Phys. Rev. B} {\bf{75}}, 104425 (2007)}.

\bibitem{Sosin} S. S. Sosin, L. A. Prozorova, A. I. Smirnov, A. I. Golov, I. B. Berkutov, O. A. Petrenko, G. Balakrishnan, and M. E. Zhitomirsky, 
\href{https://doi.org/10.1103/PhysRevB.71.094413}{{Phys. Rev. B} {\bf{71}}, 094413 (2005)}.

\bibitem{Ben3} N. Ben Amor, M. Bejar, M. Hussein, E. Dhahri, M. A. Valente, E. K. Hlil, Synthesis, Magnetic Properties,
\href{https://doi.org/10.1007/s10948-011-1344-9}{{J. Supercond. Nov. Magn.} {\bf{25}}, 1035 (2012)}.

\bibitem{Wolf} B. Wolf, U. Tutsch, S. Dorschug, C. Krellner, F. Ritter, W. Assmus, and M. Lang,
\href{https://doi.org/10.1063/1.4961708}{{J. Appl. Phys.} {\bf{120}}, 142112 (2016)}.

\bibitem{YaoDong} Y-D Wu, Q-Y Dong, Y. Ma, Y-J Ke, N. Su, X-Q Zhang, L-C Wang, Z-H Cheng,
\href{https://doi.org/10.1016/j.matlet.2017.04.007}{{Mater. Lett.} {\bf{198}}, 110 (2017)}.

\bibitem{Tkac} V. Tkac, R. Tarasenko, E. Tothova, Z. Bujnakova, K. Tibenska, A. Orendacova, V. Sechovsky, M. Orendac,
\href{https://doi.org/10.1016/j.jallcom.2019.151719}{{J. Alloys Compd.} {\bf{808}}, 151719 (2019)}.



\bibitem{Krempa} W. Witczak-Krempa, G. Chen, Y. Baek Kim, and L. Balents,
\href{http://link.aps.org/doi/10.1146/annurev-conmatphys-020911-125138}{{Annu. Rev. Condens. Matter Phys.} {\bf{5}}, 57 (2014)}.



\bibitem{Abhishek1} A. Juyal, A. Agarwal, and S. Mukhopadhyay,
\href{https://doi.org/10.1103/PhysRevB.95.125436}{Phys. Rev. B {\bf{95}}, 125436 (2017)}.
\href{https://doi.org/10.1103/PhysRevLett.120.096801}{Phys. Rev. Lett. {\bf{120}}, 096801 (2018)}.

\bibitem{Bikash} B. Ghosh, V. K. Dwivedi, and S. Mukhopadhyay,
\href{https://doi.org/10.1103/PhysRevB.102.144444}{Phys. Rev. B {\bf{102}}, 144444 (2020)}.


\bibitem{Vinod1} V. K. Dwivedi and S. Mukhopadhyay,
\href{https://doi.org/10.1063/1.5100316}{J. Appl. Phys. {\bf{125}}, 223901 (2019)}.

\bibitem{Vinod2} V. K. Dwivedi and S. Mukhopadhyay,
\href{https://doi.org/10.1063/1.5125254}{J. Appl. Phys. {\bf{126}}, 165112 (2019)}.

\bibitem{Vinod3} V. K. Dwivedi, A. Juyal and S. Mukhopadhyay,
\href{https://doi.org/10.1088/2053-1591/3/11/115020}{Mater. Res. Express {\bf{3}}, 115020 (2016)}.

\bibitem{Vinod4} V. K. Dwivedi and S. Mukhopadhyay,
\href{https://doi.org/10.1016/j.jmmm.2019.04.049}{J. Magn. Magn. Mater. {\bf{484}}, 313 (2019)}.
\href{https://doi.org/10.1016/j.physb.2019.07.006}{Physica B: Condens. Matter. {\bf{571}}, 137 (2019)}.


\bibitem{Vinod5} V. K. Dwivedi and S. Mukhopadhyay,
\href{https://doi.org/10.1063/1.4917801}{{AIP Conf. Proc.} {\bf{1665}}, 050160 (2015)}.
\href{https://doi.org/10.1063/1.4980569}{{\bf{1832}}, 090016 (2017)}.
\href{https://doi.org/10.1063/1.5033067}{{\bf{1953}}, 120002 (2018)}.
\href{https://doi.org/10.1063/1.5033100}{{\bf{1953}}, 120035 (2018)}.




\bibitem{Ding} X. Ding, B. Gao, E. Krenkel, C. Dawson, J. C. Eckert, S-W Cheong, and V. Zapf,
\href{https://doi.org/10.1103/PhysRevB.99.014438}{Phys. Rev. B {\bf{99}}, 014438 (2019)}.

\bibitem{Kazak} N. V. Kazak, M. S. Platunov, Yu. V. Knyazev, N. B. Ivanova, O. A. Bayukov, A. D. Vasiliev, L. N. Bezmaternykh, V. I. Nizhankovskii, S. Yu. Gavrilkin, K. V. Lamonova, S. G. Ovchinnikov,
\href{https://doi.org/10.1016/j.jmmm.2015.05.081}{J. Magn. Magn. Mater {\bf{393}}, 316 (2015)}.


\bibitem{Midya} A. Midya, P. Mandal, Km. Rubi, R Chen, J-S Wang, R. Mahendiran, G. Lorusso, and M. Evangelisti,
\href{https://doi.org/10.1103/PhysRevB.93.094422}{Phys. Rev. B {\bf{93}}, 094422 (2016)}.



\bibitem{Bonilla} C. M. Bonilla, J. Herrero-Albillos, F. Bartolome, L. M. Garcia, M. Parra-Borderias, and V. Franco,
\href{https://doi.org/10.1103/PhysRevB.81.224424}{Phys. Rev. B {\bf{81}}, 224424 (2010)}.



\bibitem{Das} M. Das, S. Roy, N. Khan, and P. Mandal,
\href{https://doi.org/10.1103/PhysRevB.98.104420}{Phys. Rev. B {\bf{98}}, 104420 (2018)}.

\bibitem{Paromita} P. Mukherjee and S. E. Dutton,
\href{https://doi.org/10.1002/adfm.201701950}{Adv. Funct. Mater. {\bf{27}}, 1701950 (2017)}.

\bibitem{Sanjay} S. Singh, L. Caron, S. W. D'Souza, T. Fichtner, G. Porcari, S. Fabbrici, C. Shekhar, S. Chadov, M. Solzi, and C. Felser,
\href{https://doi.org/10.1002/adma.201505571}{Adv. Mater. {\bf{28}}, 3321 (2016)}.

\bibitem{Santanu1} S. Pakhira, C. Mazumdar, R. Ranganathan, and M. Avdeev,
\href{https://doi.org/10.1038/s41598-017-07459-3}{Sci. Rep. {\bf{7}}, 7367 (2017)}.

\bibitem{Santanu2} S. Pakhira, C. Mazumdar, R. Ranganathan, S. Giri, and Maxim Avdeev,
\href{https://doi.org/10.1103/PhysRevB.94.104414}{Phys. Rev. B {\bf{94}}, 104414 (2016)}.

\bibitem{Chaudhary} V. Chaudhary and R.V. Ramanujan,
\href{https://doi.org/10.1038/srep35156}{Sci. Rep. {\bf{6}}, 35156 (2016)}.

\bibitem{Tapas} T. Samanta, I. Das, and S. Banerjee,
\href{https://doi.org/10.1063/1.2775050} {Appl. Phys. Lett. {\bf{91}}, 082511 (2007)}













					
\end{thebibliography}
\end{document}